%% file: ec2021_main.tex
\documentclass[format=acmsmall, review=false]{acmart}
\usepackage{acm-ec-21}
\usepackage[english]{babel}
\usepackage{graphicx}
\usepackage{xcolor}
\usepackage{framed}
\usepackage[normalem]{ulem}
\usepackage{enumerate}
\usepackage{algorithmic}
\usepackage[ruled, vlined,noend]{algorithm2e}
\usepackage{wrapfig} 
\usepackage{caption}
\usepackage{subcaption}
\usepackage{xcolor}
\usepackage{hyperref}
\usepackage{cleveref}
\theoremstyle{definition}

\usepackage{pdfpages}

\SetCommentSty{mycommfont}
\usepackage{amsmath}
\usepackage{bbm}
\usepackage{braket}

\setcitestyle{acmnumeric}

\title[QAOA Universal]{An upper bound on the Universality of the Quantum Approximate Optimization Algorithm}

\author{J Ceasar Aguma }

\input{sects/abstract}

\begin{document}

\begin{titlepage}

\maketitle
\end{titlepage}


\input{sects/intro}
\input{sects/proof}

\input{sects/conclusion}


\bibliographystyle{ACM-Reference-Format}
\bibliography{ref}

\end{document}

%% file: sects/abstract.tex
\begin{abstract}

Using lie algebra, this brief text provides an upper bound on the universality of QAOA. That is, we prove that the upper bound for the number of alterations of QAOA required to approximate a universal gate set is within $O(n)$. 

\end{abstract}



%% file: sects/intro.tex
\section{Introduction}
Adding to Farhi's introduction of QAOA\cite{farhi2014quantum}, Seth Lloyd presented a proof that QAOA is universal by defining new $H_z$ and $H_x$ operates that facilitate nearest neighbor interactions between qubits on a 1D lattice\cite{qaoauniv}. This is done by "turning on and off" terms in the operators using threshold wavelengths (coefficients $w_x$) on the terms of the Hamiltonian governing the system's interactions. In a follow up paper using lie algebra, a proof was given for how these operators and others can be generated from a set of {$H_z$,$H_x$}. This text attempts to answer a follow up question from the universal setting of QAOA.

\subsection{Research Question}
For a problem size $n$ and a choice of $H_z$ and $H_x$ acting on $n$ qubits
we say QAOA is universal if any element in the full unitary group $U(2^n)$ is
approximated to arbitrary precision (up to a phase) from the lie alegbra of {$H_z$, $H_x$}.
Therefore, we ask, for what sequence length p (number of alterations) is the set = {$H_z$, $H_x$} universal under QAOA dynamics.

%% file: sects/proof.tex
\section{PROOF}
\subsection{Method}
The method is quite simple, we look to at the length required to generate, Pauli-X,Y,Z operators, the CNOT on nearest neighbor qubits and the CNOT on two qubits separated by any desired length(long range CNOT). Clearly the long range CNOT would be the bottle neck as it requires the most operations or alterations. The proof for the upper bound therefore simply finds the number of alterations required to generate the long range CNOT.

\subsection{Long-range-CNOT}
This section finds the length required to approximate a CNOT on two qubits at a distance $n$ from each other on the 1-dimensional line.
\subsubsection{\textbf{Step 1 : Represent the CNOT in diagonal form}}
    \par we know that the CNOT on nearest neighbors is, $$ CNOT_{k,k+1} = \ket{0}_k\bra{0} \bigotimes I_{k+1} + \ket{1}_k\bra{1} \bigotimes X_{k+1}  $$
    If we expand the length between the two qubits $k$ and $k+d$, where $d = 0,1,2....,n$., we see that $$ CNOT_{k,k+d} = \ket{0}_k\bra{0} \bigotimes I^{\bigotimes n-1} + \ket{1}_k\bra{1} \bigotimes I^{\bigotimes n-2} \bigotimes X_{k+d}  $$ 
    which is equivalent to,
    $$ CNOT_{k,k+d} = \frac{1}{2}(I^{\bigotimes n} + Z_k \bigotimes I^{\bigotimes n-1} + I^{\bigotimes n-1} \bigotimes X_{k+d} - Z_k \bigotimes I^{\bigotimes n-2}  \bigotimes X_{k+d} )$$
    From intiutive inspection, the term $Z_k \bigotimes I^{\bigotimes n-2}  \bigotimes X_{k+d} $ is the most expensive in the $CNOT_{k,k+d}$ expression. We will therefore use this term for finding an upper bound.

\subsubsection{\textbf{Step 2 : Show that required operators in the definition of CNOT can be generated} from {$H_z$, $H_x$}}
    \par We want to show that we can generate $Z_k \bigotimes I^{\bigotimes n-2}  \bigotimes X_{k+d} $ from {$H_z$, $H_x$}.
    
    We will start by showing how to generate $Z_{k}I_{k+1}X_{k+2}$ then a show an iterative step to generate $Z_{k}I_{k+1}I_{k+2}X_{k+3}$. The same step can used to generate up to $Z_{k}I_{k+1}I_{k+2}........X_{k+n}$. 
    
    We need to generate $Z_{k}Z_{k+1}Z_{k+2}$ , $X_{k+1}Y_{k+2}$, commute the two to generate $Z_{k}I_{k+1}X_{k+2}$.
        
    From the properties of lie algebra, we know that if R and S are in the lie algebra then their commutators [R,S] and [S,R] are also in the lie algebra.
    
    if n is odd: From $H_{z}$, we can separate out the term $$H_{z2} = \gamma_{AB}H_{AB} + \gamma_{BA}H_{BA} =  \gamma_{AB} \sum_{j=1}^{\frac{n-1}{2}} Z_{2j}Z_{2j+1} + \gamma_{BA} \sum_{j=0}^{\frac{n-3}{2}} Z_{2j+1}Z_{2j+2}$$ \cite{qaoauniv1} \newline Then from $H_x$, we can generate $$ X_{even} = \sum_{j=1}^{\frac{n-1}{2}} X_{2j} $$ as shown in \cite{qaoauniv1}.
    
    Now commuting these two, $$ [H_{z2}, X_{even}] = \gamma_{AB} \sum_{j=1}^{\frac{n-1}{2}} Y_{2j}Z_{2j+1} + \gamma_{BA} \sum_{j=0}^{\frac{n-3}{2}} Z_{2j+1}Y_{2j+2} \triangleq H^{e}_{yz}$$
    
    Again commuting $$ [ H^{e}_{yz}, H_{z2}] = \gamma_{AB}^{2} \sum_{j=1}^{\frac{n-1}{2}} X_{2j} +  2\gamma_{AB}\gamma_{BA} \sum_{j=1}^{\frac{n-1}{2}} Z_{2j-1}X_{2j}Z_{2j+1}  + \gamma_{BA}^{2} \sum_{j=0}^{\frac{n-3}{2}} X_{2j+2} $$
    
    From the above we can separate out $$ H_{zxz} = 2\gamma_{AB}\gamma_{BA} \sum_{j=1}^{\frac{n-1}{2}} Z_{2j-1}X_{2j}Z_{2j+1} $$ and commute again $$ \frac{1}{2i}[[H_{zxz},X_{k+1}],Y_{k+2}] = Z_{k}I_{k+1}X_{k+2} $$ 
    
    (Note that $X_{k+1}$ and $Y_{k+2}$ can easily be generated from $H_{AB}$ and $X_{k}$ which are generated in \cite{qaoauniv1}.)
    
    With $Z_{k}I_{k+1}X_{k+2}$, we use $Y_{k+2}Z_{k+3}$ and $Z_{k+2}Y_{k+3}$ in an iterative step to generate terms up to $Z_{k}I_{k+1}I_{k+2}........X_{k+n}$.
    
    \par if n is even: From $H_{z}$, we can separate out the term $$H_{z2} = \gamma_{AB}H_{AB} + \gamma_{BA}H_{BA} =  \gamma_{AB} \sum_{j=1}^{\frac{n}{2}-1} Z_{2j}Z_{2j+1} + \gamma_{BA} \sum_{j=0}^{\frac{n}{2} - 1} Z_{2j+1}Z_{2j+2}$$ And from $H_x$, we can generate $$ X_{odd} = \sum_{j=0}^{\frac{n}{2}-1} X_{2j} $$ as shown in \cite{qaoauniv1}.
    
    Now commuting these two, $$ [H_{z2}, X_{odd}] = \gamma_{AB} \sum_{j=1}^{\frac{n}{2}-1} Z_{2j}Y_{2j+1} + \gamma_{BA} \sum_{j=0}^{\frac{n}{2}-1} Y_{2j+1}Z_{2j+2} \triangleq H^{o}_{yz}$$
    
    Again commuting $$ [ H^{o}_{yz}, H_{z2}] = \gamma_{AB}^{2} \sum_{j=1}^{\frac{n}{2}-1} X_{2j+1} +  2\gamma_{AB}\gamma_{BA} \sum_{j=1}^{\frac{n}{2}-1} Z_{2j-1}X_{2j}Z_{2j+1}  + \gamma_{BA}^{2} \sum_{j=0}^{\frac{n}{2}-1} X_{2j+1} $$ 
    
    From the above we can separate out $$ H_{zxz} = 2\gamma_{AB}\gamma_{BA} \sum_{j=1}^{\frac{n}{2}-1} Z_{2j-1}X_{2j}Z_{2j+1} $$ and commute again $$ \frac{1}{2i}[[H_{zxz},X_{k+1}],Y_{k+2}] = Z_{k}I_{k+1}X_{k+2} $$
    
    (Note that $X_{k+1}$ and $Y_{k+2}$ can easily be generated from $H_{AB}$ and $X_{k}$ which are generated in \cite{qaoauniv1}.)
    
    With $Z_{k}I_{k+1}X_{k+2}$, we use $Y_{k+2}Z_{k+3}$ and $Z_{k+2}Y_{k+3}$ in an iterative step to generate terms up to $Z_{k}I_{k+1}I_{k+2}........X_{k+n}$.
    
\subsubsection{\textbf{Step 3: find an upper bound on the sequence length required to approximate any of the operators}}
    \par From counting, we find that it takes length of $p = 3$ to generate $Z_{k}I_{k+1}X_{k+2}$ and $p \leq 10$ to get $X_{k}, Y_{k}, Z_{k} X_{k}Y_{k+1}, Z_{k}Y_{K+1}$, and $Y_{k}Z_{K+1}$. Therefore for the iterative step we need a constant length of $p \leq 12 = t$ which we repeat at most $n-2$ times. We can conclude that we need length $p = tn$ alternations which implies $O(n)$ upper bound on alternations to generate  $CNOT_{k,k+x}$ using QAOA dynamics.

%% file: sects/conclusion.tex
\section{Conclusion}
Intuitively, it is likely that the lower bound too is within $\omega(n)$ but a complete proof has not availed itself yet. If one wishes to collaborate on this idea, please reach out at jaguma@uci.edu
